\documentclass[prb,aps,showpacs,twocolumn]{revtex4}
\usepackage{graphicx}
\usepackage{dcolumn}
\usepackage{bm}
\usepackage{color}
\begin{document}
\title{Reversible enhancement of the magnetism of ultrathin Co films by H adsorption}

\author{S. Gallego\email{sgallego@icmm.csic.es}, N. Sanchez, S. Martin,  and
M.C. Mu\~noz}
\affiliation{
Instituto de Ciencia de Materiales de Madrid, Consejo Superior de
Investigaciones Cient{\'{\i}}ficas, Cantoblanco, 28049 Madrid, Spain}
\author{L. Szunyogh}
\affiliation{Department of Theoretical Physics, 
Budapest University of Technology and Economics, Budafoki \'ut 8. H1111
Budapest}

\date{\today}

\begin{abstract}
By means of {\it ab initio} calculations, we have investigated the effect of
H adsorption on the structural, electronic and
magnetic properties of ultrathin Co films on Ru(0001). Our calculations
predict that H occupies hollow
sites preserving the two-dimensional 3-fold symmetry.
The formation of a complete H overlayer leads to a very stable surface with
strong H-Co bonds. 
H tends to suppress surface features, in particular,
the enhancement of the magnetic moments of the bare film.
The H-induced effects are mostly confined to the Co atoms bonded to H,
independent of the H coverage or of the thickness and the
structure of the Co film.
However, for partial H coverages 
a significant increase occurs in the magnetic moment for the
surface Co atoms not bonded to H, leading to a net enhancement of surface
magnetism.
\end{abstract}

\maketitle

\section{Introduction}

Hydrogen is an ubiquitous element, present even in ultra-high-vacuum chambers,
although very
difficult to detect.
It shows a large sticking probability in most transition metal surfaces
\cite{christmann}, among them Co.
Adsorption of H on transition metals was intensively studied in the past
decades, mainly motivated
by the understanding of the action of residual gases in catalytic and
embrittlement processes.
Much less attention has been paid to
the influence of H on the magnetic properties, though it is known to have a
remarkable effect.
For example, the adsorption of H$_2$ molecules on the surface of magnetic
transition metals
produces a reduction of surface magnetism \cite{gradmann}. In addition,
H coverage shifts the critical
thickness for the spin reorientation transition (SRT) of Ni films \cite{robert},
and a reversible manipulation of the magnetic exchange coupling of Fe/V
superlattices can be achieved upon loading the V layers with H \cite{hjorvarsson}.

Previous experimental results characterizing the structure and energetics of H
adsorption on
different Co surfaces and nanostructures with two-dimensional (2D) 3-fold
symmetry,
as Co(0001) \cite{habermehl} or Co nanoislands on Cu(111) \cite{sicot}, indicate
that H prefers
adsorption sites preserving the symmetry of the 2D lattice.
This is in agreement with {\it ab initio} calculations of the hcp Co(0001)
termination \cite{klinke}.
Even for Co($10\overline{1}0$) the formation of superstructures leading to
adsorption sites of
3-fold symmetry has been reported \cite{ernst}.
The evaporation of molecular H$_2$ on different Co surfaces
results in the dissociative chemisorption of H atoms
up to a monolayer (ML) completion \cite{sicot}. For higher doses, molecular H$_2$
physisorption occurs.
The desorption of H from Co nanoislands on Cu(111) takes place locally and
leaves the Co surface as it was
prior the H evaporation, in
contrast to the substrate exposed areas \cite{sicot}.
This surface character of the H-Co interaction is also supported by total energy
calculations,
which show a negligible weight of the H induced states even in subsurface Co
atoms \cite{klinke,mubarak}.
In addition, no dependence of the binding energy on the H coverage is obtained \cite{klinke}.
Moreover, H subsurface sites are highly unstable, indicating a low probability
for H diffusion to the bulk.

As compared to other transition metal surfaces, there are only a few
studies of the influence of H on the
electronic and magnetic properties of Co surfaces and thin films.
Spectroscopic measurements indicate a quenching of the $sp$-like surface states
and
a shift of the $d$ surface states downward in energy upon H adsorption
\cite{greuter,himpsel}.
Moreover, the Co electronic properties can be tuned individually controlling the
amount of adsorbed H \cite{koopmans}.
The calculated density of states (DOS) also indicates that H states appear below
the $d$
valence band (VB) \cite{klinke}. When spin-polarization is included, it can be
shown
that this results in the quenching of the surface-induced enhancement of the Co
spin moment,
at least for 1 ML of H coverage \cite{mubarak}.

Recently, we studied the magnetic properties of ultrathin Co films grown on
Ru(0001) \cite{prl}.
This is a unique system showing a double SRT linked to the completion of Co
atomic layers.
The SRT is intimately related to the structural distortion of the
ultrathin Co film. Both Co and Ru
are hcp metals, but with very different lattice parameters: 2.51 \AA \ and 2.71
\AA, respectively.
While 1 ML of Co grows epitaxially, the second layer starts compressing towards
the Co lattice,
leading to an intermediate 2D lattice parameter of 2.6 \AA.
In this work we study the influence of H adsorption on the structural,
electronic and magnetic properties
of these ultrathin Co/Ru(0001) films, with special emphasis on the 2 ML
thick ones which show perpendicular magnetic
anisotropy. The study of the magnetic anisotropy will be the subject of a
forthcoming paper.
Our results indicate that H bonds strongly to the surface and tends to form a
complete
overlayer leading to a very stable surface system.
Partial H coverage, although thermodinamically stable, presents slightly smaller
energies of adsorption.
As we will show, while 1 ML of H quenches the surface magnetism, lower
H coverage may enhance it. This can be a crucial factor
regarding the role of residual H in the measured
magnetic properties of Co surfaces.

\section{Computational methods}

We have performed {\it ab initio} calculations within the local density
approximation (LDA), combining two different approaches.
First, we have performed an exhaustive search of the equilibrium adsorption
positions of H
using slab models, a plane-wave basis set and the projector-augmented wave (PAW)
method \cite{paw}
as implemented in the Vienna Ab-initio Simulation Package (VASP) \cite{vasp}.
For the most stable configurations, we have then determined
the electronic and magnetic properties within the fully-relativistic framework
of the Screened Korringa-Kohn-Rostoker
(SKKR) method \cite{skkr}, which allows for a smooth matching of the
surface region to the semiinfinite bulk and lacks of the quantum size effects
inherent to slab models.

For the calculations with VASP, an energy cutoff of 350 eV has been used. The
structures
are modeled by periodically repeated slabs of 6 to 10 Ru layers covered by 2 Co MLs on one side,
and adding 1/4 ML or 1 ML of H on top of or underneath the outermost Co
layer.
A vacuum region of 12 \AA \ is left between both slab surfaces
which prevents any spurious interaction between them.
This model also guarantees the recovery of the bulk properties at the inner
Ru layers of the slab.
The search for the most stable H adsorption sites has been performed starting
from different
H geometries and allowing full relaxation of the atomic positions of H, Co and
the
outermost Ru layer. A uniform sampling of the k-mesh
centered in $\Gamma$  has been used,
and convergence for the most stable structures has been obtained
using k-samplings of up to ($15 \times 15 \times 1$) for $(1 \times 1)$
geometries
and ($8 \times 8 \times 1$) for the $(2 \times 2)$ unit cells corresponding to
adsorption of 0.25 ML of H.
Under these conditions, convergence in
total energy differences below 1 meV is achieved.

By using the optimized geometry with the most
stable H adsorption sites,
the detailed electronic structure and the magnetic
moments were calculated in terms of the
fully relativistic SKKR method.\cite{skkr}
The SKKR formalism provides, in a natural way, layer-resolved physical
quantities such as charges
and magnetic moments.
It, thus, manifests a perfect tool for the analysis of
the H-Co interactions, and of the influence of the Co film thickness
in the H induced properties. For this purpose, we have extended our calculations
to films of 3 to 6 Co MLs.
Details of the SKKR calculations for the Co/Ru system can be found
elsewhere \cite{prl,njp}.
The H atom is modeled by using an atomic sphere of radius 0.59 \AA,
placed at the equilibrium distance to Co determined by VASP. 

\section{Structural properties}

\begin{figure}[ht]
\begin{center}
\includegraphics[angle=270,width=0.8\columnwidth,bb=116 144 396 612,clip]{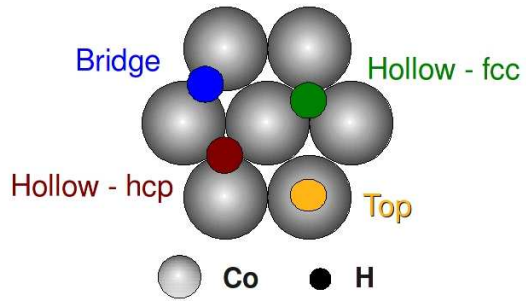}
\caption{(Color online) High symmetry sites for the adsorption of H on hcp Co.}
\label{Hposit}
\end{center}
\end{figure}

The stable adsorption site of H on 2Co/Ru(0001) has been explored starting
from different geometries, depicted in figure \ref{Hposit}: 
top, bridge, hollow (either fcc or hcp) and off-symmetry positions,
both at surface and subsurface sites.
In addition, for 1 ML coverage, the influence of structural
modifications present in the Co
film has been considered, such as the variation of the 2D lattice parameter, or
the existence
of stacking faults which alter the Co/Ru stacking sequence from the hcp AB/AB to
BC/AB \cite{jorge}.

A summary of the results is compiled in Table \ref{tstruc}.
In all cases, the most stable adsorption site corresponds to H at fcc hollow
surface positions.
The table shows the energy difference with respect to the equilibrium configuration
only for the most stable structures.
The rest of geometries considered, including on-top positions, provide total
energies larger by more than 1 eV. Our results are in good agreement
with previous ones relative to different Co surfaces and thin films
\cite{klinke,mubarak}:
H tends to stick at the surface trying to preserve the 3-fold symmetry of the Co
lattice,
independent of the detailed Co film structure or the H coverage.
The energy difference between adsorption sites provides an estimate of the
diffusion barriers
for H in the surface. Although the trends are similar for all surfaces studied,
partial H coverage or
alteration of the stacking sequence at the Co film may modify the diffusion
barriers significantly.
Regarding subsurface sites, the barrier for the diffusion of H into the Co film
decreases as the H coverage increases, but it is always large enough to support
the low probability of the penetration of H into the film, in good agreement with the 
experiments of thermal desorption through formation of H$_2$ \cite{sicot}.

The results indicate that the adsorption of H is governed by the local
surface formation of strong H-Co bonds.
Supporting this, for all structures considered the H-Co distance remains around
1.7 \AA, independent
of the H coverage and the Co film structure.
The largest variations of the H-Co distance are found to depend on the
H adsorption site,
where H is coordinated to a different number of Co atoms:
from 1.5 \AA \ for the on-top surface adsorption (coordination 1), to 1.9 \AA \
for hollow subsurface sites (coordination 6).

\begin{table}[htbp]
\caption{Energy (in meV) per H atom with respect to the most stable
fcc hollow site for different H adsorption sites on 2 ML Co on Ru(0001),
both for 1 ML and 0.25 ML H coverage. The dependence on the
in-plane lattice parameter a$_{2D}$ and the stacking sequence
of the Co film are shown for the high H coverage.}
\renewcommand{\arraystretch}{1.2} 
\begin{tabular}{cp{2mm}ccp{3mm}ccc}
\hline \hline
 & & \multicolumn{2}{c}{Co film} & &\multicolumn{3}{c}{H adsorbate} \\
   & & a$_{2D}$ (\AA ) & stacking & & hcp   &  bridge  &    subsurf. \\
\hline
  1 ML & & 2.7 & AB &  & 27   &  289 &     $> 420$ \\
       & &      & BC &  & 24   & 708 &      $> 430$ \\
\hline
       & & 2.6 & AB &  & 27   &  278 &     $> 480$ \\
       & &      & BC & & 108   & 708  &     $> 550$ \\
\hline
 0.25 ML  & & 2.6 & AB & & 45 & 236 & $>1000$ \\
\hline \hline
\end{tabular}
\label{tstruc}
\end{table}

A further insight can be obtained comparing
the energetics and structure of the equilibrium configurations of the Co film 
covered by different amounts of H. From now on we will restrict our
considerations to the experimental value of a$_{2D}$, 2.6 \AA.
The adsorption energy per H atom is defined as:
\begin{equation}
E_{ads} = - {1\over N_H} [E_{HCoRu} - E_{CoRu} - N_H E_H]
\end{equation}
where $N_H$ refers to the number of adsorbed H atoms per unit cell, $E_{HCoRu}$
and $E_{CoRu}$ respectively to the total energies of the slabs with and without
adsorbed H and $E_H$ to the energy of isolated H. The adsorption
energies for the different
H coverages are reported in Table \ref{eners}. Their positive sign indicates
that adsorption is favorable in all cases. When comparing with the formation
energy of
the H$_2$ molecule, 2.44 eV/H atom, we can see that both H coverages are
highly stable against desorption, in good agreement with the experimental
evidence and the previous theoretical calculations for similar Co surfaces
\cite{klinke,mubarak}.
An alternative approach to determine the stability of the H covered surfaces
comes from the evaluation of the work function, $\Phi$, also provided in Table
\ref{eners} as
obtained with the SKKR method. The increase of $\Phi$ with H coverage, specially
for the complete H overlayer, corroborates the large sticking coefficient of H
at Co
surfaces even for ultrathin films, and is in good agreement with the results
obtained
for Co films on Cu(001) \cite{mubarak}.

\begin{table}[htb]
\caption{Adsorption energy and work function for 2Co/Ru(0001), either bare or
covered by H,
together with the variation of the Co-Co and Co-Ru interlayer
spacings relative to the unrelaxed
Ru bulk, $\Delta$d$_{CC}$ and $\Delta$d$_{CR}$, respectively.
The two values of $\Delta$d$_{CC}$ for 0.25 ML coverage
correspond to Co atoms unbonded/bonded to H.}
\label{eners}
\renewcommand{\arraystretch}{1.2} 
\begin{tabular}{ccccc}
\hline \hline
H coverage  &    E$_{ads}$ (eV)   &  $\Phi$ (eV)    & $\Delta$d$_{CC}$ (\AA) &
$\Delta$d$_{CR}$
(\AA)  \\
\hline
Bare      &      &  6.03  &   -0.29 &    -0.06 \\
 0.25 ML  & 3.34 &  6.09  &   -0.30/-0.27 & -0.07  \\
 1 ML     & 3.41 &  7.02  &   -0.22 & -0.09   \\
\hline
\end{tabular}
\end{table}

Table \ref{eners} also provides the variation of the interlayer
spacings in the Co film.
As expected from the reduced atomic volume of Co compared to Ru, even
for the bare surface
there is a significant contraction of the Co-Co interlayer distance, d$_{CC}$,
trying to compensate the in-plane expansion. Accordingly, the Co-Ru distance,
d$_{CR}$,
is only slightly reduced. The large volume compensation effect
suppresses the
oscillatory relaxation of interlayer distances typical of transition metal
surfaces,
although an oscillatory behavior can be identified in the layer-resolved
electronic properties for larger Co thicknesses.
Upon H adsorption, the progressive filling of free bonds attenuates the
surface induced
relaxations, but still the volume compensation remains. Thus, the
resulting relaxation pattern is different from that of
other H covered films where Co retains its 2D lattice constant \cite{mubarak}.
In spite of this difference, we can conclude that H always acts as an attenuator
of surface effects.
It is worth noticing the slight corrugation of 0.03 \AA \ at
the Co surface layer for 1/4 ML H coverage, driven by
the Co atoms bonded and unbonded to H. Remarkably,
the surface induced contraction of d$_{CC}$ is enhanced for those unbonded.

\section{Electronic and magnetic properties}

\begin{figure*}[th]
\begin{center}
\includegraphics[angle=270,width=1.8\columnwidth,clip]{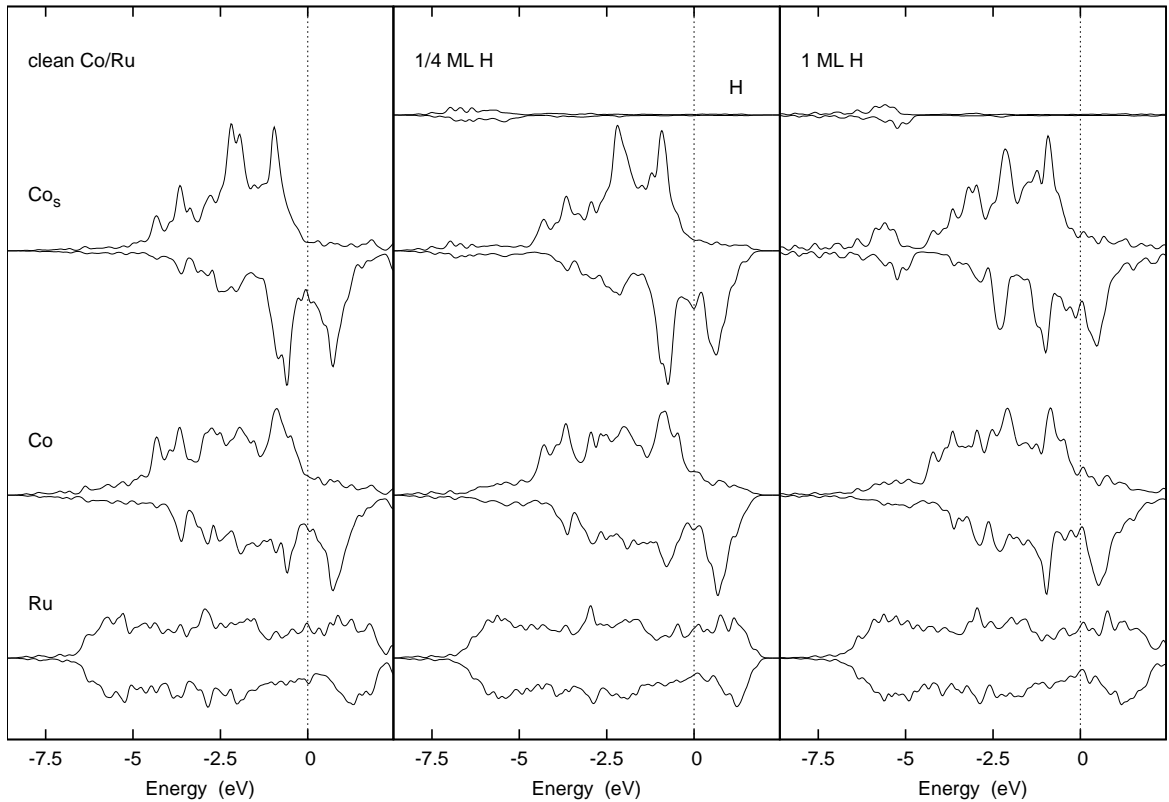}
\caption{DOS projected on the H, surface Co (Co$_s$), interface
Co and Ru atoms of the 2Co/Ru(0001) surface system either bare (left) or covered 
by 1/4 ML (middle) and 1 ML (right) of H. The Co$_s$ DOS for 1/4 ML H coverage
corresponds to Co atoms bonded to H. Energies are related to the Fermi level.}
\label{dos}
\end{center}
\end{figure*}

Next we discuss the layer-resolved electronic and magnetic properties of the
2 Co/Ru(0001) film with and without H coverage.
The atom projected DOS and magnetic moments are shown in Fig.
\ref{dos}  and in Table \ref{mom}, respectively.
The DOS of the bare film shows the narrowing induced by the reduced coordination
at the surface plane.
In the case of Co, with the majority spin band almost full, this causes an
increase of the
magnetic moment. As H adsorbs, the surface effects start to
attenuate, and, upon completion of 1 ML of H, the magnetization of the
two Co layers becomes almost uniform. The H-states localize at the bottom
of the $d$ valence band (VB), and show a significant spin polarization induced
by the hybridization with Co, even though the net magnetic moment of H is almost zero.
In turn, H-induced features can be identified in a region dominated by $sp$ states for the
surface Co atoms bonded to H. The weight of these features for the rest of Co atoms,
including those at the surface layer for 0.25 ML of H (not shown in the figure),
is negligible. 

The short range of the H-induced effects can also be inferred from
the magnetic moments of the surface Co atoms bonded and not bonded to H for
0.25 ML coverage, see Table \ref{mom}.
Noticeably, the partial H coverage results in a significant increase of the
magnetic moments of the Co atoms unbonded to H.
This is evidenced both in the fully relativistic calculations performed with the
SKKR method and
in those performed with slab models. Though the arbitrary assignment of the
atomic radii affects the actual values of the magnetic moments for each atom,
the relative differences between the structures calculated 
under the same conditions provide a
reliable physical picture. Furthermore, the total magnetic moment per
slab, which is not subject to the assignment of the individual atomic radii, also
supports the enhancement of the surface magnetism for partial H coverages: 9.64
$\mu_B$ (bare),
10.88 $\mu_B$ (0.25 ML) and 9.72 $\mu_B$ (1 ML).
This enhancement can be understood in terms of the strong local H-Co
interaction, which limits the Co-Co interactions within the surface plane.
Quite interestingly, similar trends for the orbital moments of the
Co atoms upon H dosing can be inferred from Table \ref{mom}. From this we expect
that the adsorption of H 
affects the magnetic anisotropy energy of the Co/Ru(0001) system in a
non-trivial manner, as well. 

\begin{table}[htbp]
\caption{Layer resolved magnetic moments obtained with slab models
(m$_{slab}$), as well as spin (m$_S$) and orbital (m$_L$) contributions to the 
magnetic moments calculated
by the relativistic SKKR method for the 2Co/Ru(0001) surface system
covered by different amounts of H. 
The moments are given in $\mu_B$ and the labels of the atoms are the
same as in Fig. \ref{dos}. 
For 0.25 ML H coverage, the moments of the Co atoms
unbonded/bonded to H are distinguished.}
\renewcommand{\arraystretch}{1.2} 
\renewcommand{\tabcolsep}{0.4pc} 
\begin{tabular}{cccccc}
\hline \hline
          &    Ru     &  Co &  Co$_S$ & H     & \\
\hline
m$_{slab}$ & -0.05   &  1.31  & 1.69   &  --  & bare \\
           &  -0.07  &  1.30/1.23 & 1.76/1.54 & -0.02 & 0.25 ML \\
           &  -0.02  &  1.27  &  1.27  & -0.01 & 1 ML \\
\hline
m$_S$      & -0.03  &  1.51  &  1.75  & -- &
bare\\
           & -0.01  &  1.56/1.50 & 1.94/1.75 & 0.04 & 0.25 ML \\
           &    -0.01  &  1.52  &  1.58  &  0.01 & 1 ML  \\
\hline
m$_L$     & 0.00  &  0.08  &  0.11  & -- & bare
\\
          & 0.00  &  0.08  &  0.13/0.10 & 0.00 & 0.25 ML \\
          & 0.00  &  0.08  &  0.07  &  0.00 & 1 ML \\
\hline \hline
\end{tabular}
\label{mom}
\end{table}

\begin{figure}[htb]
\begin{center}
\includegraphics[width=\columnwidth,clip]{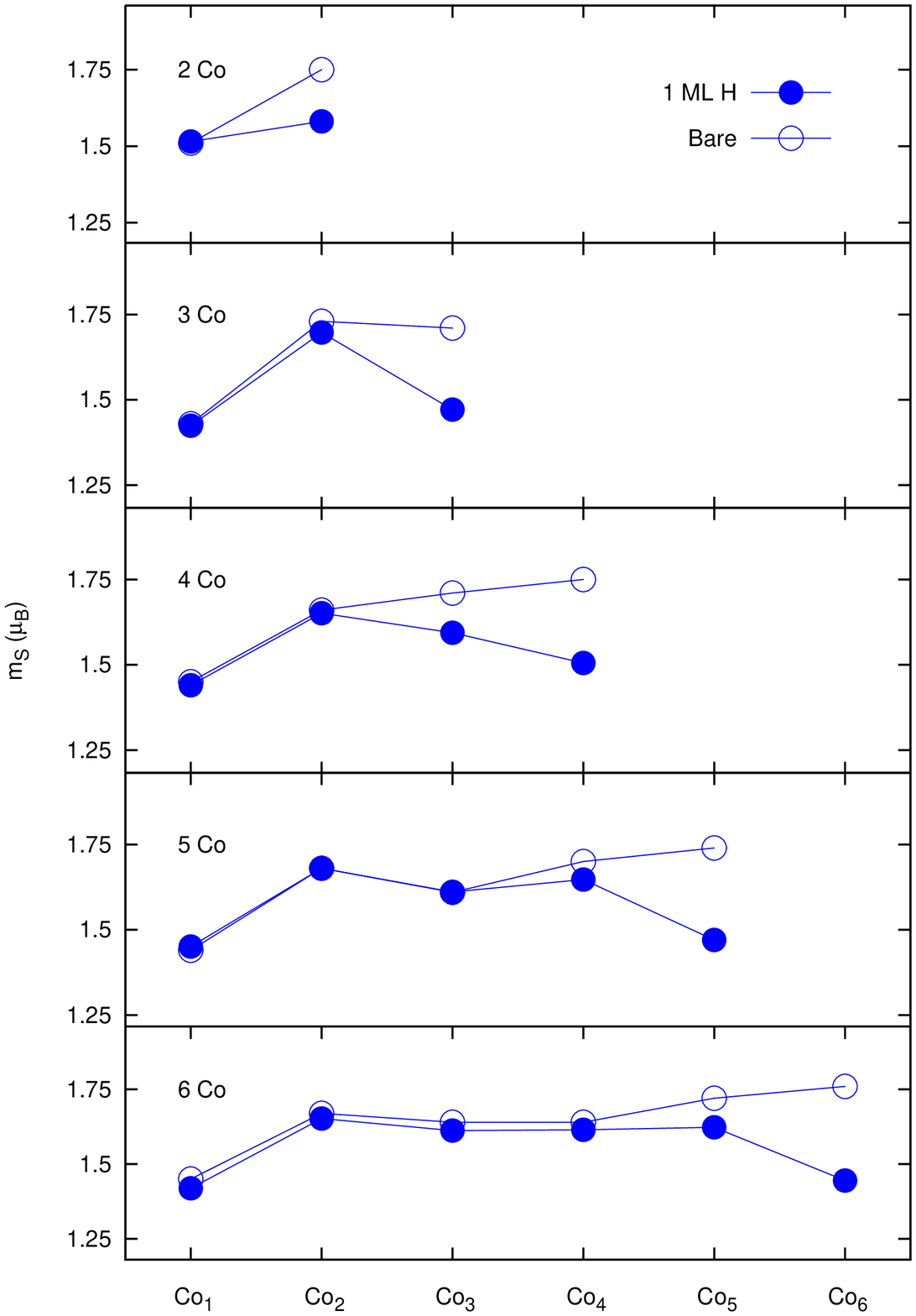}
\caption{\label{fmoms} (Color online) Layer-resolved Co spin magnetic moment for Co films of
2 to 6 MLs (from top to bottom), either bare (empty circles) or covered by H (solid circles).
Layers are numbered from the Ru interface.}
\end{center}
\end{figure}

\begin{figure}[hbt]
\begin{center}
\includegraphics[width=\columnwidth,clip]{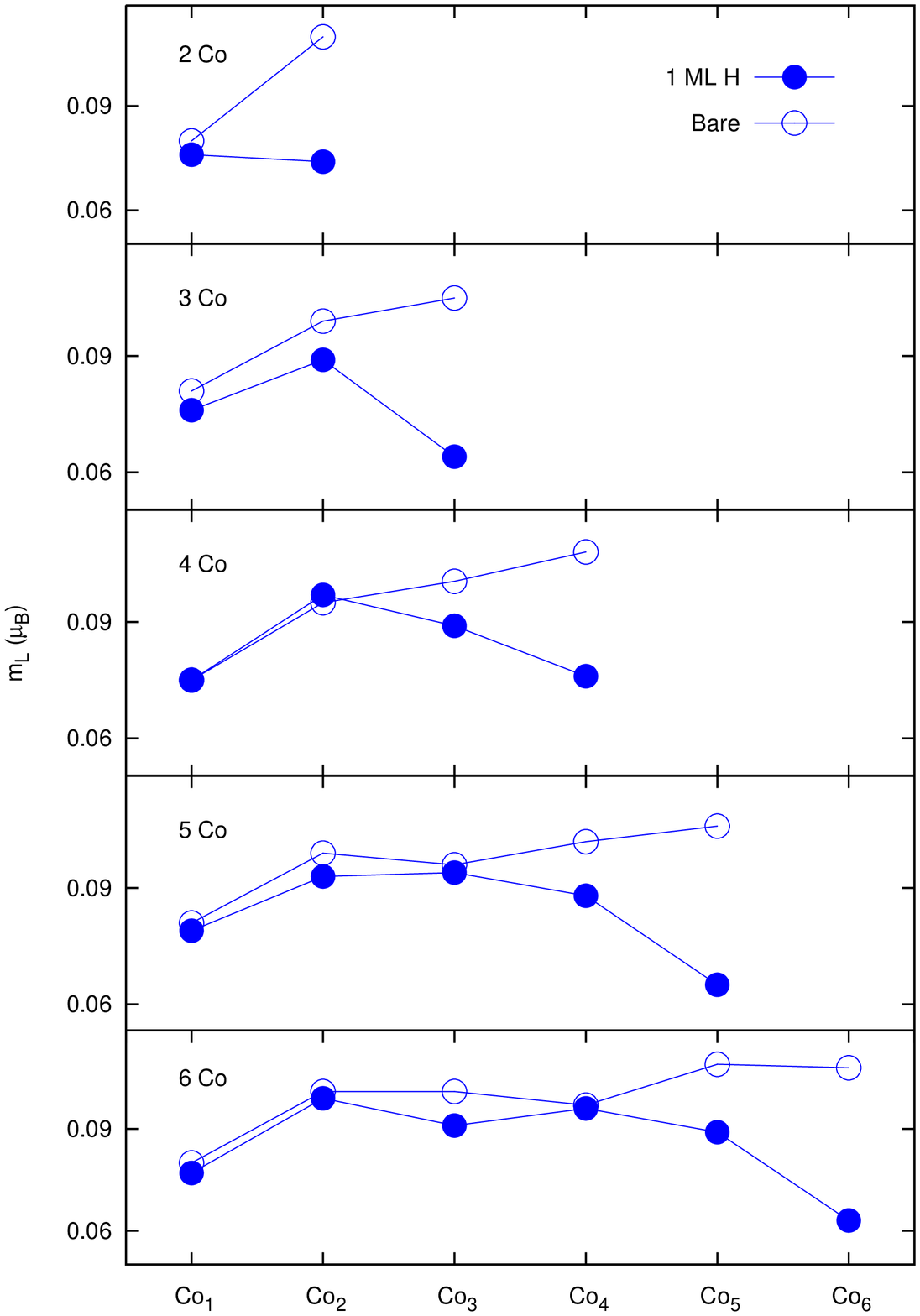}
\caption{\label{fmoml} (Color online) Same as figure \protect\ref{fmoms} for the orbital contributions to the magnetic moment.}
\end{center}
\end{figure}

Finally, we discuss the H induced effects for thicker Co films. 
To this end, we performed SKKR calculations for Co films of up
to 6 MLs covered by 1 ML of H. 
Figures \ref{fmoms} and \ref{fmoml} show respectively the layer resolved
values of the spin (m$_S$) and orbital (m$_L$) contributions to the
magnetic moment for the different Co thicknesses, together with those at the
corresponding bare surfaces.
Upon H adsorption there is always a strong reduction of both m$_S$ and m$_L$
at the surface Co atoms. On average, the reduction is
around 0.25 $\mu_B$  for m$_S$ and 0.04 $\mu_B$ for m$_L$.
By contrast, at the subsurface Co layer the reduction of the magnetic moment
is low, becoming negligible for the layers underneath.
Although the specific values of m$_S$ and m$_L$ at each layer
involve in a complex way the oscillatory layerwise variation of the electronic 
properties, the effect of H does not depend on the Co thickness. 
Furthermore, it is very similar for both components of the magnetic moment.

Not shown in the figure, in all cases the H atom acquires a m$_S$ of 0.01 $\mu_B$, and
no orbital polarization. Similarly, the components of the moment induced at the interface 
Ru atom are around -0.02 $\mu_B$ (m$_S$) and -0.001 $\mu_B$ (m$_L$) for all Co thicknesses,
without any significant variation as compared to the bare film.
The layer resolved DOS (not shown here) are consistent with this
conclusion, with H induced features
almost restricted to the surface Co layer in each case.
Thus we can conclude that the H induced effects are essentially restricted to
the surface and independent of the Co thickness.

\section{Conclusions}

In terms of {\it ab initio} calculations, we performed a detailed theoretical analysis 
of the stability, electronic structure and magnetism of Co/Ru(0001) films upon H 
adsorption. Our main observations are as follows:
H adsorbs on Co at the hollow sites by preserving the threefold symmetry of the
system. The adsorption is endothermic for all H coverages, with the largest binding
energy corresponding to the complete H overlayer. Furthermore, the probability of
H desorption or segregation is low.
Concerning the structural, electronic and magnetic properties, the effect of H
has a local surface character and is almost independent of the detailed Co
thickness and structure. H always attenuates surface effects.
However, for partial H coverages, this leads to the enhancement of the surface 
induced features for Co atoms not bonded to H, and in particular to an enhancement
of the net surface magnetic moment.
Our results point to the importance of residual H in magnetic measurements, and 
to the possibility for
the reversible manipulation of the properties of ultrathin Co
films and surfaces by tuning H adsorption.

\section{Acknowledgments}

This work was supported by the Spanish Ministry of Science and Technology
under contracts MAT2006-05122, HH2006-0027 and MAT2009-14578-C03-03, 
and by the Hungarian Research
Foundation under contracts  OTKA K68312 and K77771.

\end{document}